\documentclass[twocolumn,secnumarabic,amssymb, nobibnotes, aps, prd]{revtex4-1}

\usepackage{graphicx} % support the \includegraphics command and options

\begin{document}
\title{Efficient Fluorescence Collection from Trapped Ions with an Integrated Spherical Mirror}
\author{G. Shu}
\email{shugang@u.washington.edu}
\author{N. Kurz}
\author{M.R. Dietrich}
\author{B.B. Blinov}
\affiliation{Department of Physics, University of Washington, Seattle WA 98195-1560}
\date{\today}
\begin{abstract}
Efficient collection of fluorescence from trapped ions is crucial for quantum optics and quantum computing applications, specifically, for qubit state detection and in generating single photons for ion-photon and remote ion entanglement. In a typical setup, only a few per cent of ion fluorescence is intercepted by the aperture of the imaging optics. We employ a simple metallic spherical mirror integrated with a linear Paul ion trap to achieve photon collection efficiency of at least $10\%$ from a single Ba$^+$ ion. An aspheric corrector is used to reduce the aberrations caused by the mirror and achieve high image quality.
\end{abstract}
\maketitle

With their long coherence time and the straightforward manipulation of internal and external states to store and process quantum information, trapped ions are among the most promising candidates for practical quantum computation~\cite{Blinov-Hperfine-2004,Blatt-Ca-2004,Blatt-QCIon-2008}. Coupling of ionic qubits to single photons through spontaneous emission offers an attractive alternative~\cite{Moehring-PhotonAtomNetwork-2007} to the more traditional ion trap quantum computing architectures~\cite{Cirac-Zoller-1995,Kielpinski-2002}. Reliable interconversion between quantum states of single ions and single photons thus becomes one of the most crucial tasks. A robust ion-photon coupling scheme is indispensable for implementing remote quantum gates between ions~\cite{Duan-RemoteGate-2006,Monroe-2009} and quantum repeaters~\cite{Duan-Repeater-2004}. Multi-element refractive optical systems are most widely used for ion and atom fluorescence collection and imaging~\cite{Weiss-AtomImag-2007, Greiner-2009,Thompson-CaPenningTrap-2004,Blatt-AtomImageInterference-2001,Schlosser-dipoletrap-2001,Sortais2007,Tey-2008}. Recently, such custom-designed, in-vacuum lenses capable of collecting about 4\% of the photons from an ion were used to achieve 0.2\% fiber coupling efficiency for 397~nm photons from Ca$^+$~\cite{Almendros-2009}. High-finesse optical cavities have also been used~\cite{Keller-IonCavity-2004,Drewsen-CloudinCavity-2008,blat-cavity-2009} but their full potential has yet to be attained with ions where close proximity of the dielectric cavity mirrors is detrimental to the trapping itself. 

Simple reflective optics offer an attractive alternative and have been previously implemented in non-imaging fluorescence detectors such as~\cite{Bergmann1979}. Compared to their complex refractive counterparts, reflective optics can make large deflection of light propagation with considerably fewer elements and simpler surfaces. Because no transparency is necessary, metallic optical surfaces can be used, which can be placed in close proximity to the ion without affecting the trap performance. Much higher numerical apertures (N.A.) can thus be achieved with comparatively small size optics. Though placing an ion in the focus of a parabolic mirror may be the best scenario~\cite{Leuchs-FreespaceModeConvert-2007}, and traps designed for such implementation have been successfully demonstrated~\cite{Wineland-stylustrap-2009}, a high-N.A. parabola is a complicated surface to fabricate, especially in a small-scale device, which would be required for a scalable trapped-ion quantum information processor. Spherical mirrors are much simpler and can be fabricated using standard microelectromechanical systems (MEMS) technology~\cite{Kraft-2005}. The intrinsic large image aberrations caused by the spherical mirror can be compensated with optics located outside the vacuum chamber.  

In this article we demonstrate and characterize the performance of a spherical mirror with an effective N.A. of at least 0.6, limited only by the linear quarupole trap geometry. We design and implement an aspheric corrector element to reduce the aberrations caused by the spherical mirror. We also suggest a straightforward modification of the trap design to further improve both the image quality and the collection efficiency.

Our trap is derived from a working linear quadrupole trap design originally developed by the Monroe group~\cite{Boris-BroadbandCooling-2006}. We modify the trap by attaching a concave spherical aluminum-coated mirror to its frame (Figure~1). The mirror has a curvature radius of 20~mm, and is ground to $19\times 19$~mm$^2$ square from its original 25~mm diameter to fit in the frame opening. A precision-machined holder supports the mirror from below and ensures that its focus is aligned with the trap center. The aluminum surface of the mirror is electrically connected to the trap ground to avoid any residual charge accumulation. 

\begin{figure}[htp]
\centering
\includegraphics[width=0.4\textwidth]{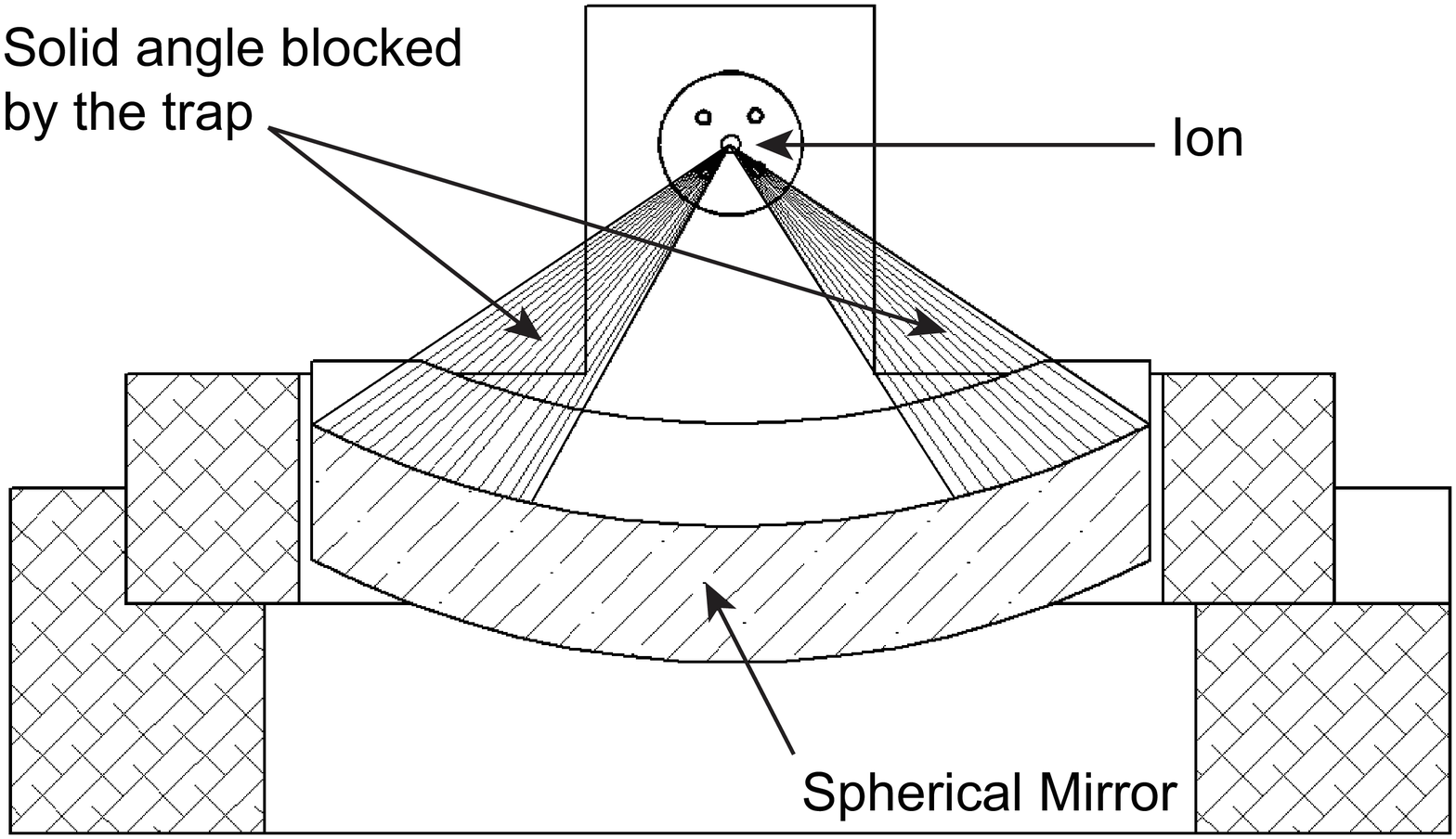}
\caption{Cross section of the linear quadrupole trap with the integrated mirror, viewed along the axis of the trap. The photons from the ion are partially blocked by the trap quadrupole rods.}
\end{figure}

From geometric calculation, we find the mirror to have a N.A. of 0.82, which gives 2.72~sr or 21.6\% of the entire 4$\pi$ solid angle. However, due to our linear trap structure, the two quadrupole rods located between the ion and the mirror effectively block 1.27~sr of this solid angle (see Figure 1). After photons are reflected, another $4.3\%$ is scattered by the trap electrodes. Taking both into account we arrive at an effective N.A. of 0.63 (corresponding to a solid angle of approximately 1.39~sr or 11\%). A more detailed description of the system and its assembly procedure can be found in~\cite{Shu-2009}. The trap/mirror assembly is placed in an ultrahigh vacuum system (typical pressure $5\times10^{-11}$~torr) with ample optical access for laser beams and ion fluorescence collection. The ions are Doppler-cooled using 493~nm and 650~nm lasers~\cite{Kurz-2008, Matt-2009}. With 60~V DC applied to the endcap electrodes and $\sim$5~W RF signal at 22.5~MHz applied to the quadrupole, the typical trap secular frequencies are $2\pi\times200$~kHz (axial) and $2\pi\times1$~MHz (radial); trapping times of several days have been observed.

In our initial tests described in~\cite{Shu-2009}, ion images formed directly by the mirror were demonstrated albeit with very large aberrations. To increase the image quality, and to improve the light collection efficiency, we design and implement an aspheric corrector element similar in function to the Schmidt plate~\cite{PofOptics}. The very large N.A. of our system requires the corrector to have a customized shape, which is numerically calculated. To calculate the corrector surface profile we employ a segment-fitting algorithm, that is, we divide the entire surface into many segments and apply Snell's law on each segment. When the number of segments increases, they can be joined to form a smooth curve. Note that this is different from our earlier algorithm for deriving the corrector shape~\cite{Shu-2009}, where the curve is generated with asymptotic integration. Compared with the old one, the new method is simpler and more effective; the corrector curves generated by the old and the new method are plotted in Figure~2(a). The new numerically derived curve can be satisfactorily fit to a 10$^{th}$ order polynomial for optical simulations. Remarkably, our simulations show that the spherical mirror + corrector system is nearly equivalent in performance to a perfect parabolic mirror (Figure~2(b,c)). For comparison, the performance of the corrector curve derived by asymptotic integration is included in Figure~2(b,c), which is about two orders of magnitude worse.

\begin{figure*}[htp]
\centering
\includegraphics[width=0.8\textwidth]{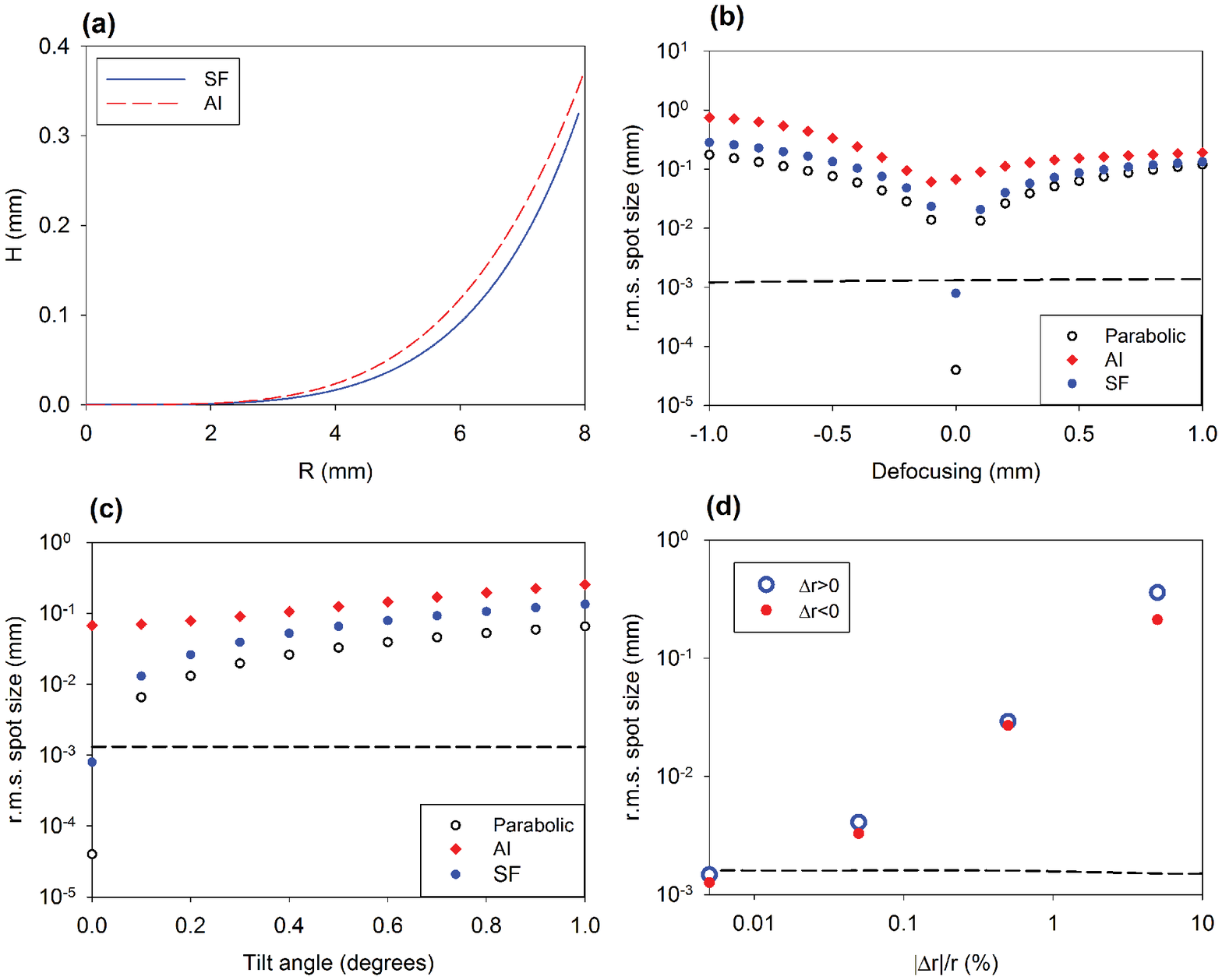}
\caption{Simulations of corrector performance (Color online). (a) The curves generated by the segment-fitting algorithm (SF) and the asymptotic integration (AI), with corrector thickness $H$ plotted as a function of distance from the optical axis $R$. (b) and (c) Simulated performance of a spherical mirror with the two corrector elements as compared to a parabolic mirror performance for defocussing (b) and the optical axis tilt (c).  Dashed line indicates the diffraction limit. Diffraction was not taken into account while calculating the RMS spot sizes; points below the dashed line would result in a diffraction-limited image. (d) Image size degradation due to the deviation of the spherical mirror radius of curvature $r$ from the value used in calculating the corrector shape. Dashed line indicates the diffraction limit.}
\end{figure*}

To test the corrector performance, we machine in-house several correctors out of clear acrylic plastic to approximately 25~$\mu$m precision using a computer numerical controlled (CNC) lathe, and hand-polish them to 0.1~$\mu$m surface roughness to achieve suitable optical quality. The corrector is designed to make rays coming from the ion parallel; to form an image, a 4-element, diffraction-limited, 0.1 N.A. microscope objective plus a singlet lens are used. The image is formed on an Electron-Multiplying CCD (EMCCD) camera. Figure~3 shows the images and the image profiles of (the same) single $^{138}$Ba$^+$ ion without (a and b) and with (c and d) the corrector. Without the corrector, only the paraxial light is focused into a small spot, while the majority of light is spread out over a large area; therefore the spot is much dimmer. Its asymmetric shape comes from a combination of the large spherical aberration, higher order aberrations, and the coma due to small misalignment of the ion position with respect to the mirror focus.  With the corrector the image shape and brightness are substantially improved. It has a much smaller r.m.s. spot size, although the full width at half maximum (FWHM) of the small bright spot profile is similar to that without the corrector. The system can resolve ions $10~\mu$m apart, though the image is still far from the diffraction limit, which is a prerequisite for efficient single-mode fiber coupling. We can estimate the fiber coupling efficiency by calculating the overlap between the mode profile of a single mode optical fiber (Nufern 460HP fiber, mode field diameter 3.5$\mu$m) and the observed ion image profiles. Without corrector, the efficiency would be approximately $0.4\%$, while with the corrector it would increase to about $0.8\%$. Both values are very far from ideal, but the gain due to the corrector is obvious. 

Several factors could have contributed to the worse than expected performance of our system. First, the ion location misalignment with respect to the optical axis and the focus of the mirror (see Figure~2(b,c)) would generate aberrations and increase the image size. Second, the corrector calculated for the 20~mm mirror would not work so well for a mirror with a different radius of curvature. The effect of this deviation is shown in Figure~2(d). The radius of curvature of the spherical mirror used in our system has a tolerance of $5\%$, and the observed image spot size can be easily explained by even a fraction of this deviation. Furthermore, due to our fabrication procedure the corrector may deviate from the calculated shape by as much as $25~\mu$m. Theoretically, if both the mirror and the corrector are fabricated and aligned perfectly, the ion image should reach the diffraction limit.

\begin{figure*}[htp]
\centering
\includegraphics[width=0.8\textwidth]{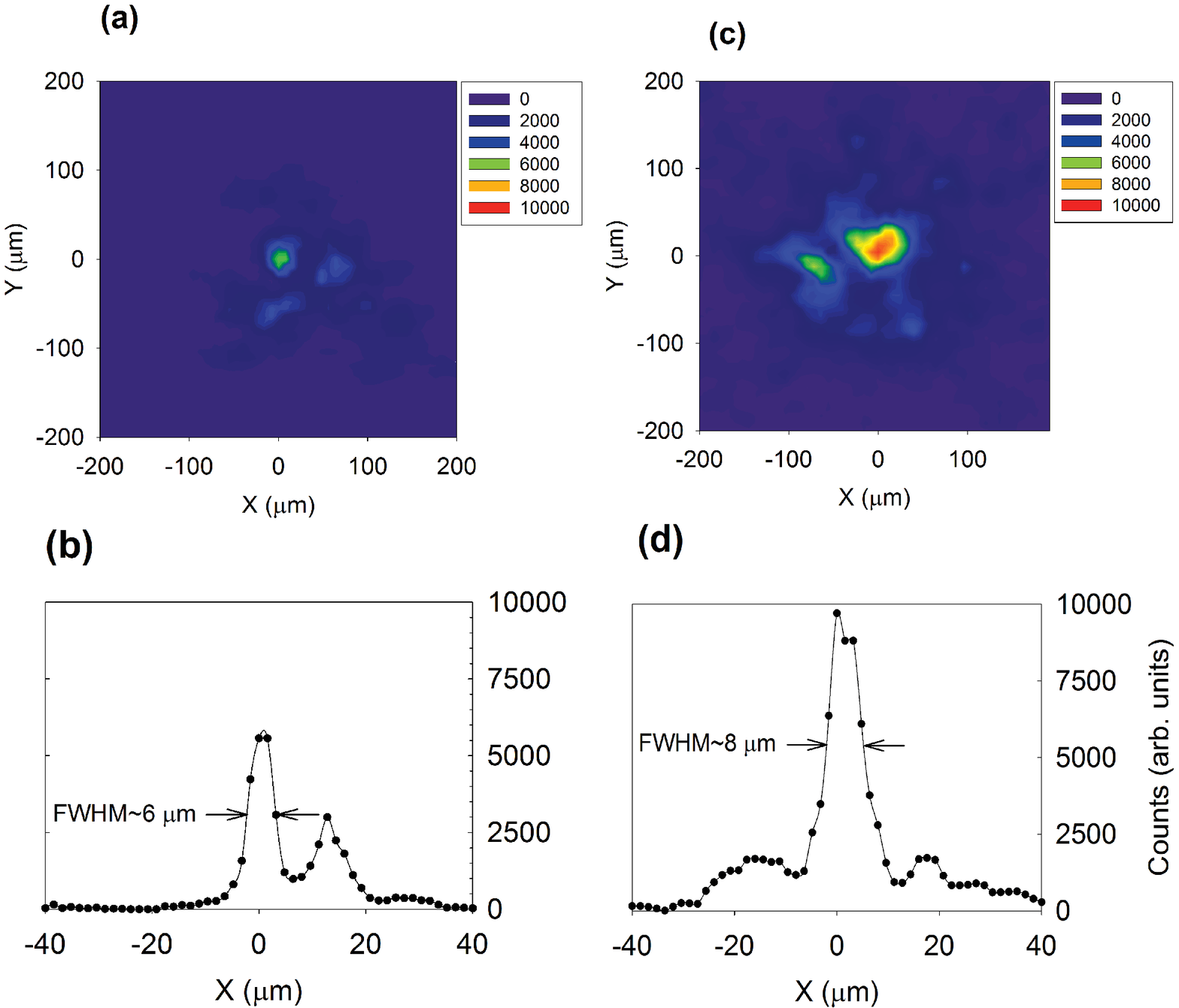}
\caption{Single ion images formed by the spherical mirror without (a) and with (c) the corrector (Color online). The corresponding zoom-in intensity profile plots (Y=0) are shown in (b) and (d).}
\end{figure*}

To measure the effective N.A. of the mirror and to quantify the performance improvement due to the corrector, we use a single photon source based on CW laser excitation of a single Ba$^+$ ion, similar to that described in~\cite{Almendros-2009}. The energy level diagram of $^{138}$Ba$^+$ and the laser pulse sequence used in the experiment are shown in Figure~4. The 493~nm single photons are generated by first optically pumping the ion to the metastable $5D_{3/2}$ state (life time $\sim80$~s) by exposing the ion to 493~nm laser for 300~ns. The 493~nm laser is then turned off and the 650~nm repumping laser is turned on for 500~ns to optically pump the ion back into the ground $6S_{1/2}$ state, generating a single 493~nm photon. We confirm that no more than one photon is generated by observing lack of photon coincidence at zero time delay in a $g^{(2)}$  measurement~\cite{Kurz-2009}. The photons are detected by the single-photon-counting photomultiplier tubes (PMTs). The cycle is repeated $10^6$ times for each optical setup to achieve sufficient statistical precision.

\begin{figure*}[htp]
\centering
\includegraphics[width=0.8\textwidth]{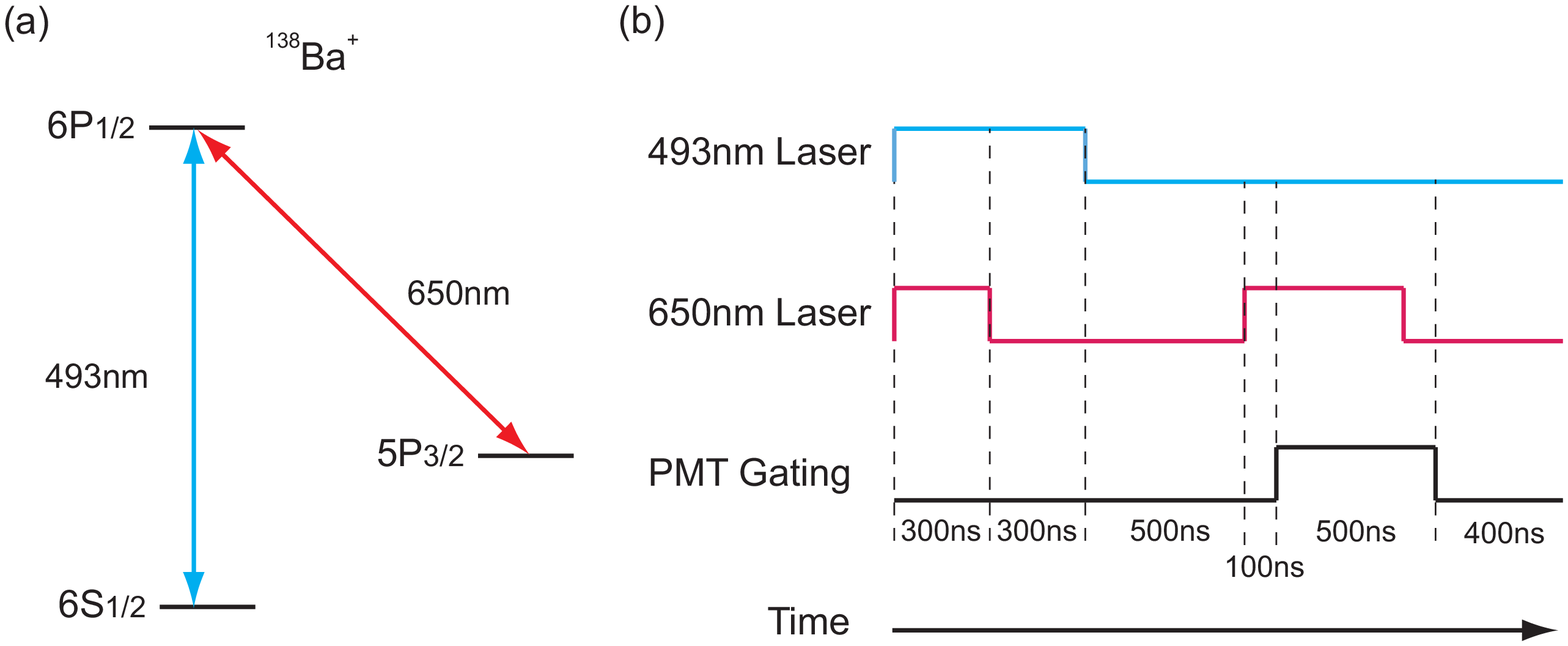}
\caption{$^{138}$Ba$^+$ energy level diagram (a) and single photon generation logic sequence (b)(Color online). The ion is cooled for 300~ns with both laser on. Then it is exposed to 493~nm laser for 300~ns to be pumped to $5D_{3/2}$ state. A 500~ns exposure to 650~nm laser excites the ion to $6P_{1/2}$ state and generates a single 493~nm photon when the ion decays back into ground state. A 500~ns time delay is inserted between the laser pulses to ensure the previous laser pulse is completely extinguished. The PMT is gated to count only during the 650~nm laser pulse. A 100~ns time delay between the start of the 650~nm laser pulse and PMT gate is necessary to compensate for the electronic signal delays.}
\end{figure*}

Next, we confirm that a single photon is indeed emitted by the ion \textit{every time}. Note that the $g^{(2)}$ experiment only proves that no more than one photon is emitted; it does not prove that there is always exactly one. We achieve this by measuring the photon counting rates for a series of known numerical apertures formed by a calibrated optic~\cite{Jungsang}. By calculating the ratio between counts and the N.A., we can estimate the source intensity. For this purpose, we use the diffraction-limited microscope objective that \textit{directly} images the ion (i.e. not using light reflected by the mirror) with a calibrated iris placed in front. In this setup the PMT photon counts can be expressed as:
\begin{equation}
C=C_0+\eta\Omega I=C_0+\eta\Omega (I_s+I_{bg})
\end{equation}
where $C_0$ is the PMT's dark counts ($\sim60$ per 1 million excitations), $\eta$ is the overall efficiency to detect single photons from the ion, $\Omega$ is the fractional solid angle of photon collection, $I_{s}$ is the intensity of the single photon source in photons per experimental cycle, and $I_{bg}$ is the count due to background light.  For each individual aperture setup, we take two measurements: one with the 650~nm laser on and one with the laser off, respectively. Single 493~nm photons are generated only when the red laser is on. By subtracting the signal measured with the 650~nm laser off from the signal with the laser on, both the dark count and the background are canceled and we get the photon count coming from the ion: 
\begin{equation}
C_{ion}=\eta\Omega I_s=\eta_{t}\eta_{pmt}\Omega I_s
\end{equation}
The overall efficiency $\eta$ is the product of $\eta_{t}$, the optical system throughput, and $\eta_{pmt}$, the quantum efficiency of the PMT ($12.7\pm1.0\%$). With five antireflection (AR) coated BK7 lenses (approximately 0.6\% loss per surface), one AR coated fused silica view port (1.0\% loss per surface) and one 492~nm interference filter (53.1\% loss), we have $\eta_{t}=0.432$ and $\eta=0.065$. Using these numbers and the measured PMT counts for different N.A. values, we find that $I_s=0.976+0.024-0.106$, which is consistent with unity, as expected. The error is dominated by the uncertainty of the PMT quantum efficiency.

We then repeat the experiment using the mirror with the corrector; the 4-element microscope objective is now used again as it was used for the imaging, but the photons are directed to the PMT. We measure $4350\pm91$ single photons per $10^6$ excitation cycles (statistical uncertainty), corresponding to about 0.43\% raw photon detection efficiency, or 3.4\% throughput to the PMT. Using Equation~2 and taking into account additional losses (aluminum mirror reflectivity loss of 9\% and the uncoated corrector throughput loss of about 8.9\%) we find an effective solid angle of $1.24\pm 0.13$~sr (combining statistical and equipment uncertainty). Without the corrector, the count is $3981\pm74$ photons per $10^6$ cycles, and the effective solid angle is $1.02\pm0.11$~sr (note that only the mirror reflectivity loss is included here). Remarkably, even with the extra loss introduced by the uncoated corrector there is a noticeable improvement in the photon count rate with the corrector. This is due to the fact that a significant fraction of ion fluorescence lies outside the $8\times20~$mm$^2$ aperture of the PMT when no corrector is used.

To eliminate the photon loss due to trap structure and to achieve image quality necessary for efficient single-mode fiber coupling, we are developing a novel trap design. We plan to use the metallic surface of the mirror itself as one of the trap electrodes, with a movable needle passing through the mirror vertex forming another electrode. By reducing the mirror radius of curvature to approximately 1/4 of the current mirror, the r.m.s. spot size will be reduced proportionally. By moving the needle with a micrometer stage, we can position the ion with a precision of a few microns to ensure proper aberration correction. We expect that this design will at least double the solid angle of the present system, and the image quality will reach the diffraction limit.

In conclusion, we use a high-N.A. spherical mirror in combination with an aspheric corrector to efficiently collect fluorescence and image a single trapped ion. The measured performance of the system is consistent with the geometric calculation. We thus conclude that small, reflective spherical optics can be a viable tool in trapped ion quantum information processing. Based on these findings we plan to build an improved trap with a smaller size, larger N.A. mirror.

We would like to thank Nathan Pegram for his help with the early parts of this work and Jungsang Kim for helpful discussion. This research was supported by the National Science Foundation Grants No. 0758025 and No. 0904004, the Army Research Office, and the University of Washington Royalty Research Fund.

%Merlin.mbs v4.21 2009-07-09.
%

%\bibliographystyle{apsrev4-1}
%\bibliography{cites}

\end{document}